\documentclass[sn-nature, pdflatex]{sn-jnl}

\usepackage{graphicx}%
\usepackage{multirow}%
\usepackage{amsmath,amssymb,amsfonts}%
\usepackage{amsthm}%
\usepackage{mathrsfs}%
\usepackage[title]{appendix}%
\usepackage{xcolor}%
\usepackage{textcomp}%
\usepackage{manyfoot}%
\usepackage{booktabs}%
\usepackage{algorithm}%
\usepackage{algorithmicx}%
\usepackage{algpseudocode}%
\usepackage{listings}%
\usepackage{natbib}

\newcommand{\degree}{\mbox{$^{\circ}$}}               
\newcommand{\micron}{\mbox{\,${\mu}$m}}               
\newcommand{\Msolar}{\mbox{\,$M_{\odot}$\/}}          
\newcommand{\Mjup}{\mbox{\,$M_{\rm Jup}$\/}}          
\newcommand{\HII}{\mbox{H\,{\footnotesize II}}}       
\newcommand{\water}{\mbox{H$_2$O}}                    
\newcommand{\methane}{\mbox{CH$_4$}}                  
\newcommand{\Teff}{\mbox{$T_{\mbox{\footnotesize eff}}$}} 
\newcommand{\oversim}[2]{\lower0.5ex\vbox{\baselineskip=0pt\lineskip=0.2ex
     \ialign{$\mathsurround=0pt #1\hfil##\hfil$\crcr#2\crcr\sim\crcr}}}
\newcommand{\eg}{\mbox{\hbox{e.g.,}}}             
\newcommand{\idest}{\mbox{\hbox{\it i.e.,}}}          
\newcommand{\cf}{\mbox{\hbox{\it cf.}}}               

\begin{document}

\title[Article Title]{Jupiter Mass Binary Objects in the Trapezium Cluster}


\author*[1]{\fnm{Samuel G} \sur{Pearson}}\email{samuel.pearson@esa.int}

\author*[1]{\fnm{Mark J} \sur{McCaughrean}}\email{mjm@esa.int}

\affil*[1]{\orgdiv{European Space Research and Technology Centre (ESTEC)}, \orgname{European Space Agency (ESA)}, \orgaddress{\street{Keplerlaan 1}, \city{Noordwĳk}, \postcode{2201 AZ}, \state{Zuid Holland}, \country{The Netherlands}}}







\abstract{A key outstanding question in star and planet formation is how far the initial mass function of stars and sub-stellar objects extends, and whether or not there is a cut-off at the very lowest masses. Isolated objects in the planetary-mass domain below 13 Jupiter masses, where not even deuterium can fuse, are very challenging to observe as these objects are inherently faint. Nearby star-forming regions provide the best opportunity to search for them though: while they are young, they are still relatively warm and luminous at infrared wavelengths. Previous surveys have discovered a handful of such sources down to 3--5 Jupiter masses, around the minimum mass limit established for formation via the fragmentation of molecular clouds, but does the mass function extend further? In a new James Webb Space Telescope near-infrared survey of the inner Orion Nebula and Trapezium Cluster, we have discovered and characterised a sample of 540 planetary-mass candidates with masses down to 0.6 Jupiter masses, demonstrating that there is indeed no sharp cut-off in the mass function. Furthermore, we find that 9\% of the planetary-mass objects are in wide binaries, a result that is highly unexpected and which challenges current theories of both star and planet formation.}




\maketitle

\section{Main}\label{sec1}
%
The Orion Nebula is arguably the most famous and well-studied \HII{} region in the sky. It is the nearest site of recent massive star formation, producing stars spanning the full spectral range from massive O-types to M dwarfs, a rich population of sub-stellar brown dwarfs, and many planetary-mass objects. Collectively, these objects are known as the Orion Nebula Cluster and the densest inner core, within 0.5 parsec of the eponymous Trapezium stars, is called the Trapezium Cluster, with a core density reaching $5\times 10^4$ stars pc$^{-3}$ \citep{McCaughrean1994}. Due to its large population of $\sim 2000$ members \citep{Morales2011}, young age (0.5--2\,Myr) \citep{Hillenbrand1997}, low foreground extinction (${\rm A}_{\rm v} \sim 1$) \citep{Scandariato2011}, and close proximity to the Sun \citep[$390\pm 2$\,pc;][]{apellaniz2022}, the Trapezium Cluster provides an ideal laboratory for studies of star and planet formation \citep{Muench2008, Odell2008}.

Sub-stellar objects below the hydrogen-burning limit \citep[0.075\Msolar;][]{Baraffe2015, dupuy2017, Chabrier2023} never reach the main sequence and continually cool, becoming fainter as they age. However, when young, sub-stellar sources remain relatively luminous and easy to detect as they shed gravitational energy while contracting: brown dwarfs also undergo a period of deuterium fusion, while sources below 13\Mjup{}, the planetary-mass objects (henceforth PMOs) do not. The Trapezium Cluster is a particularly advantageous location to study such sources: it is young and has a large enough sample size for robust population statistics, while its relative proximity, location out of the galactic plane, and the dense molecular cloud behind it help minimise contamination due to foreground or background field stars. 

Past ground- and space-based surveys of the Trapezium Cluster have revealed a rich population of brown dwarfs and PMOs down to $\sim 3$\Mjup{} \citep{Lucas2000, McCaughrean2002, Slesnick2004, Meeus2005, Lucas2005, Riddick2007, Weights2009, Anderson2011, DaRio2012, Robberto2020}, but reaching masses below that is challenging, partly because lower-mass objects are cooler and thus emit most of their energy in the thermal infrared, and partly due to the bright background of the Orion Nebula. Similarly, spectroscopically-confirmed objects below the deuterium burning limit remain relatively rare due to their faintness \citep{Lucas2001, Meeus2005, Lucas2006, scholz2012, Ramirez2012, gagne2017, Lodieu2021, Bouy2022}.

As a large, diffraction-limited, cryogenic space telescope, however, the JWST is ideally suited to pushing further into the planetary-mass domain than previously possible, and the wide range of filters allow us to search for tell-tale atmospheric features which can help distinguish between bona fide PMOs and distant field stars. Imaging surveys with NIRCam over wide areas can discover many new candidates, while multi-object follow-up spectroscopy is possible with NIRSpec.

An $11\times 7.5$ arcminute (or $1.2\times 0.8$ parsec) region of the inner Orion Nebula and Trapezium Cluster was observed using the Near Infrared Camera (NIRCam) on the NASA/ESA/CSA James Webb Space Telescope (JWST), as part of Cycle 1 GTO programme 1256\footnote{
Please refer to the following links for full resolution colour images:
\begin{itemize}
\item {\tt https://www.esa.int/Science\_Exploration/Space\_Science/Webb/\\ 
           Webb\_s\_wide-angle\_view\_of\_the\_Orion\_Nebula\_is\_released\_in\_ESASky}
\item {\tt https://www.esa.int/ESA\_Multimedia/Images/2023/09/\\
           Orion\_Nebula\_in\_NIRCam\_short-wavelength\_channel}
\item {\tt https://www.esa.int/ESA\_Multimedia/Images/2023/09/\\
           Orion\_Nebula\_in\_NIRCam\_long-wavelength\_channel}
\item {\tt https://sky.esa.int/?jwst\_image=webb\_orionnebula\_shortwave}
\item {\tt https://sky.esa.int/?jwst\_image=webb\_orionnebula\_longwave}
\end{itemize}}. A total of 34.9 hours of observing were carried out between 26 September and 2 October 2022, split across 12 filters: F115W, F140M, F162M, F182M, F187N, F212N, F277W, F300M, F335M, F360M, F444W, and F470N (see McCaughrean
\& Pearson 2023, submitted, for full details)

\begin{figure}[H]
\centering
\includegraphics[width=1\textwidth]{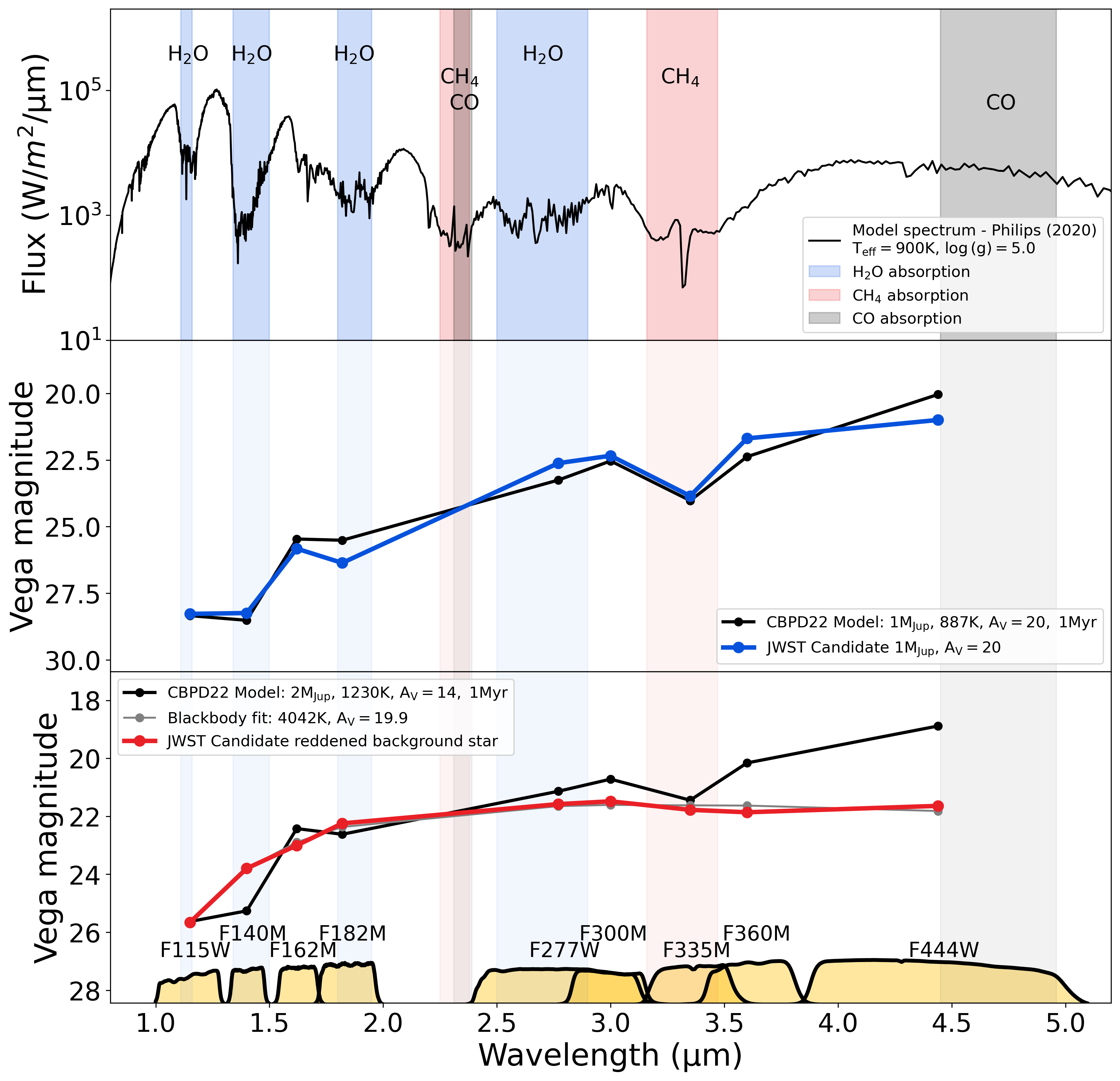}
\caption{The upper panel shows a model spectrum of a young PMO with \Teff{} = 900\,K and $\log({\rm g})=5.0$ from the ATMO\,2020 chemical equilibrium model set \cite{Philips2020}. The molecular absorption bands with line intensities greater than $5\times 10^{-23}$ cm$^{-1}$ / molecule cm$^{-2})$ at ${\rm T}_{\rm ref}$ = 900\,K for \water, \methane, and CO are shown in blue, red and black, respectively \cite{HITRAN}. Due the low temperatures of PMOs, these molecules are present in their atmospheres and absorption radically alters the spectral energy distribution into a series of narrow peaks. Using medium- and wide-band photometry, these peaks and troughs can be readily identified and provide a robust method for distinguishing PMOs from more massive background objects. In the middle panel, the black line shows synthetic photometry of a 1\,Myr, 1\Mjup{} PMO, using the atmospheric models from ATMO\,2020 \cite{Philips2020} combined with the new equation of state from Chabrier \& Debras (2021) \citep{Chabrier2021, Chabrier2023}. The model photometry has been reddened by ${\rm A}_{\rm V} = 20$. The blue line shows our JWST NIRCam photometry of a candidate $\sim 1$\Mjup{} PMO: the nominal errorbars are smaller than the markers. The strong molecular absorption dips can clearly be seen in the F140M and F182M filters due to the presence of \water, as well as a strong dip in F335M due to \methane. The match between the data and model SEDs is excellent.
The bottom panel shows synthetic photometry of a 1\,Myr, 2\Mjup{} PMO with ${\rm A}_{\rm V} = 14$ in black, alongside  NIRCam photometry of a candidate reddened background star shown in red. The candidate reddened background star does not show any molecular absorption features, and instead has much smoother spectral energy distribution: it is well fit by  blackbody with \Teff{} = 4042\,K and reddening of ${\rm A}_{\rm V} = 19.9$. For reference, the bandpasses of the nine NIRCam filters (F115W, F140M, F162M, F182M, F277W, F300M, F335M, F360M, and F444W) used to classify the sources is shown along the bottom of the plot.}
\label{figSED}
\end{figure}

Young (1\,Myr) PMOs with masses between 1--13\Mjup{} have effective temperatures of 890--2520\,K \citep{Philips2020}, which means that their spectral energy distributions (SED) peak in the range 1--3.3\micron. These SEDs are not blackbodies, but are dominated by broad molecular absorption features as seen in Figure~\ref{figSED}. The upper panel shows a model spectrum of a young PMO with \Teff{} = 900\,K and $\log({\rm g})=5.0$, taken from the ATMO\,2020 chemical equilibrium model set \citep{Philips2020}. The molecular absorption bands due \water, \methane, and CO are shown in blue, red, and black, respectively, and are seen to radically alter the SED, confining the spectrum to a series of narrow peaks and troughs. Our selection of NIRCam filters was designed to target these peaks and troughs, in order to robustly distinguish PMOs from more massive and reddened background objects. We have used photometry in the F115W, F140M, F162M, F182M, and F227W filters to measure the depth of the 1.4 and 1.9\micron{} \water{} absorption features, classifying sources according to Equation~\ref{h2o}. We have also quantified the level of \water{} absorption using the W-index, defined in Equation~\ref{h2o2}. As this index utilises the short-wavelength filters, it is susceptible to reddening and so can only be treated as a reliable indicator for low-extinction sources. To identify lower-mass, cooler sources, we use F300M, F335M, and F360M photometry to measure the 3.35\micron{} \methane{} absorption feature, classifying sources according to Equation~\ref{ch4}.

The power of medium-band near-infrared photometry to identify PMOs using these absorption features is demonstrated in Figure~\ref{figSED}. The blue curve in the middle panel shows JWST NIRCam photometry of a candidate PMO, while the black curve shows synthetic photometry derived from evolutionary models of a 1\,Myr old, 1\Mjup{} object at the distance of Orion \citep{Chabrier2023}, calculated using a new equation-of-state for dense hydrogen-helium mixtures \citep{Chabrier2021}, combined with the atmospheric models from ATMO\,2020 \cite{Philips2020}. The model fluxes have been adjusted with the further addition of ${\rm A}_{\rm V} = 20$ of reddening to best fit the candidate PMO\@. The strong molecular absorption dips can clearly be seen in the F140M and F182M filters due to the presence of \water, as well as a strong dip in F335M due to \methane. In contrast, the black curve in the bottom panel shows synthetic photometry of a model 1\,Myr, 2\Mjup{} object adjusted by ${\rm A}_{\rm V} = 14$. This clearly provides a bad fit to the JWST photometry of a candidate reddened background star, seen in red: a much better solution to its smooth SED is arrived at by assuming a reddened blackbody at \Teff{} = 4042\,K, with reddening of ${\rm A}_{\rm V} = 19.9$.

For all unsaturated stars in our JWST survey (Pearson \& McCaughrean 2023, in prep), we have fit the medium- and wide-band filter SED to evolutionary models by varying the mass and extinction, assuming a constant age of 1\,Myr and a distance of 390\,pc. We use three grids of models, one using equilibrium chemistry (CEQ), and two using non-equilibrium chemistry (NEQ$_{\rm weak}$ \& NEQ$_{\rm strong}$) \citep{Chabrier2023}. These models cover PMOs with masses in the range 0.0004\Msolar{} to 0.015\Msolar{} (0.42\Mjup to 15.7\Mjup). We also use the models which cover brown dwarfs and low mass stars from 0.01\Msolar{} to 1.4\Msolar \citep{Baraffe2015}. The extinction is allowed to vary from ${\rm A}_{\rm V}$ = 1--100. For each combination of model mass and extinction, we reddened the model SED using the reddening law from \citep{mathis} with ${\rm R}_{\rm V} = 3.1$, and calculated the $\chi^2$ goodness of fit. The lowest $\chi^2$ value was taken as the best fit. This process was also repeated using a blackbody model with \Teff{} = 500--50,000\,K and ${\rm A}_{\rm V}$ = 0--100.

\begin{figure}[H]
\centering
\includegraphics[width=1\textwidth]{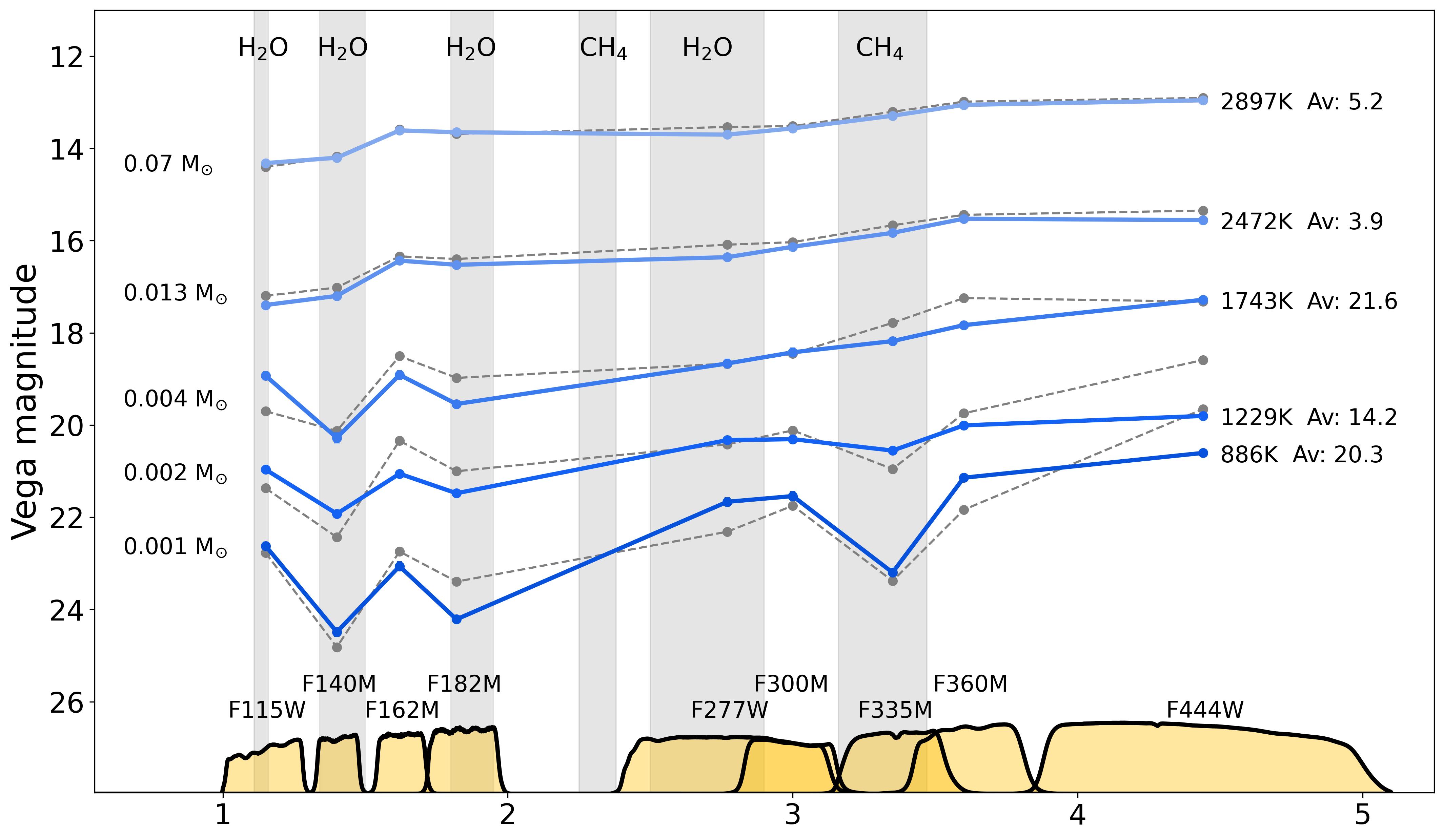}
\caption{The blue lines show dereddened NIRCam photometry for a sample of candidate brown dwarfs and PMOs in the Trapezium Cluster. This plot demonstrates how the SED sampled by our nine NIRCam medium- and wide-band filters evolves with decreasing mass for young brown dwarfs and PMOs. The light grey lines show the best-fitting unreddened model assuming an age of 1\,Myr and distance of 390\,pc \citep{Chabrier2023, Baraffe2015}. The \water{} absorption features at 1.4 and 1.9\micron{} are present for brown dwarfs and strengthen with decreasing mass, while below 5\Mjup, \methane{} absorption at 3.35\micron{} becomes prominent and also strengthens with decreasing effective temperature.}
\label{figSpectralSeries}
\end{figure}

Figure~\ref{figSpectralSeries} shows dereddened NIRCam photometry for a sample of candidate brown dwarfs and PMOs in blue. In each case, the light grey curve shows the best-fitting unreddened model assuming an age of 1\,Myr and a distance of 390\,pc \citep{Chabrier2023, Baraffe2015}. This plot demonstrates how the SED sampled by our nine NIRCam medium- and wide-band filters evolves with decreasing mass. The \water{} absorption features at 1.4 and 1.9\micron{} are already present for brown dwarfs and strengthen with decreasing effective temperature. The \methane{} absorption at 3.35\micron{} emerges at temperatures below \Teff{} $\sim$ 1500\,K, making it sensitive to PMOs below 5\Mjup. It also strengthens as the effective temperature decreases.

\begin{gather}
    {\rm F140M} - {\rm F162M} \geq 1.605({\rm F162M} - {\rm F182M}) + 0.565 \label{h2o} \\
    {\rm W index} = {\rm F115W} + 2\times({\rm F140M} - {\rm F162M} + {\rm F182M}) - {\rm F277W} \label{h2o2} \\
    {\rm F335M} - {\rm F360M} \geq 0.488({\rm F300M} - {\rm F355M}) + 0.206 \label{ch4}
\end{gather}

Background field stars are ruled out on the basis of their smooth SED, but older, cooler, foreground brown dwarfs could be identified as false positives. However, given the relative proximity of the Orion Nebula and its location out of the galactic plane, such contamination is expected to be minimal \citep{Scholz2022}. Another potential contaminant is distant galaxies. Towards the centre of the region where the extinction due to the background OMC-1 core is high \citep[${\rm A}_{\rm V} \geq$ 50--100,][]{Castets1990}, this is not a concern, but at the edges of our survey, a population of galaxies is evident (McCaughrean \& Pearson 2023, submitted). The aperture photometry technique described in the Supplementary Material detects extended sources like galaxes and, inter alia, circumstellar disks and outflow nebulosities, but also close binaries, so such sources are marked and checked manually.

In this way, we have identified 540 candidate PMOs in the Trapezium Cluster with SEDs best fit by evolutionary models with masses of 13\Mjup{} or lower. For low extinction sources (${\rm A}_{\rm V} < 10$), we also include candidates which show \water{} absorption according to Equation~\ref{h2o} and which have a W-index of $\geq$ 0.47 (Equation~\ref{h2o2}). A total of 168 of these PMO candidates also show \methane{} absorption and are best fit by models with masses of 5\Mjup{} or less. The most extreme candidate PMO in our sample has a mass of 0.6\Mjup{} or 2 Saturn masses. Our PMO candidates show a smooth continuation of the IMF to low masses, with no evidence for a sharp cutoff. We find no evidence for a large population of marginally-detected sources, which indicates that it is very unlikely that the mass function rises significantly below 1\Mjup{}. We cautiously note a moderate increase in the number of objects in the 1--3\Mjup{} range, that might be consistent with an over density of low-mass PMOs formed through ejection \citep{vanElteren2019, Scholz2022}. However, as these are currently unconfirmed PMO candidates and the masses are not well constrained, we will leave a full analysis of the IMF to future work, which will greatly benefit from scheduled follow up spectroscopy (JWST cycle 2 programme 2770).

We find that the chemical equilibrium models (CEQ) give the best fit to the data for PMOs down to 2\Mjup. Below this mass, we see that the NEQ$_{\rm weak}$ models are preferred, with CEQ second best, and NEQ$_{\rm strong}$ worst. For the 0.6--2\Mjup{} PMOs, we find that the F360M, F162M, and F444W filters that generally most badly fit by the CEQ models. This could be an indication that vertical mixing is affecting the nitrogen and carbon non-equilibrium chemistry \citep[\cf{}][]{Zahnle2014}, causing an increased abundance and absorption of CO and ${\rm CO}_2$ suppressing the flux in the 3.5--5\micron{} range. We will obtain R$\sim$100 NIRSpec prism spectroscopy for many of these PMO candidates as part of JWST cycle 2 programme 2770, which should help further investigate this tentative finding.


A remarkable finding is that a significant fraction of our candidate PMOs are in binaries. Across the initial mass function, the multiplicity fraction, defined as the fraction of primaries that have at least one companion, is seen to decrease with mass. For massive O- and B-type stars, the multiplicity fraction is close to 100\%; this fraction decreases to 50--60\% for solar type stars \citep{Duqennoy1991}, drops to 15\% for higher-mass brown dwarfs \citep[50--80\Mjup,][]{Burgasser2007}, and falls further to 8\% for lower-mass brown dwarfs \citep[20--60\Mjup,][]{Fontanive2018}. Following this trend, and also in accordance with theoretical expectations \citep[\eg][]{Bate2019}, it would be expected that the multiplicity fraction for PMOs below 13\Mjup{} should be close to zero. However, within our sample of 540 PMO candidates there are a 40 systems that have a binary companion within 1 arcsec (390\,au), and two visual triple systems, a multiplicity fraction of at least 9\%. The existence of these ``Jupiter-Mass Binary Objects'', henceforth JuMBOs, is an unexpected result that breaks a trend that holds for over three orders of magnitude in mass \citep{Fontanive2018, Fontanive2023}.

\begin{figure}[H]%
\centering
\includegraphics[width=0.9\textwidth]{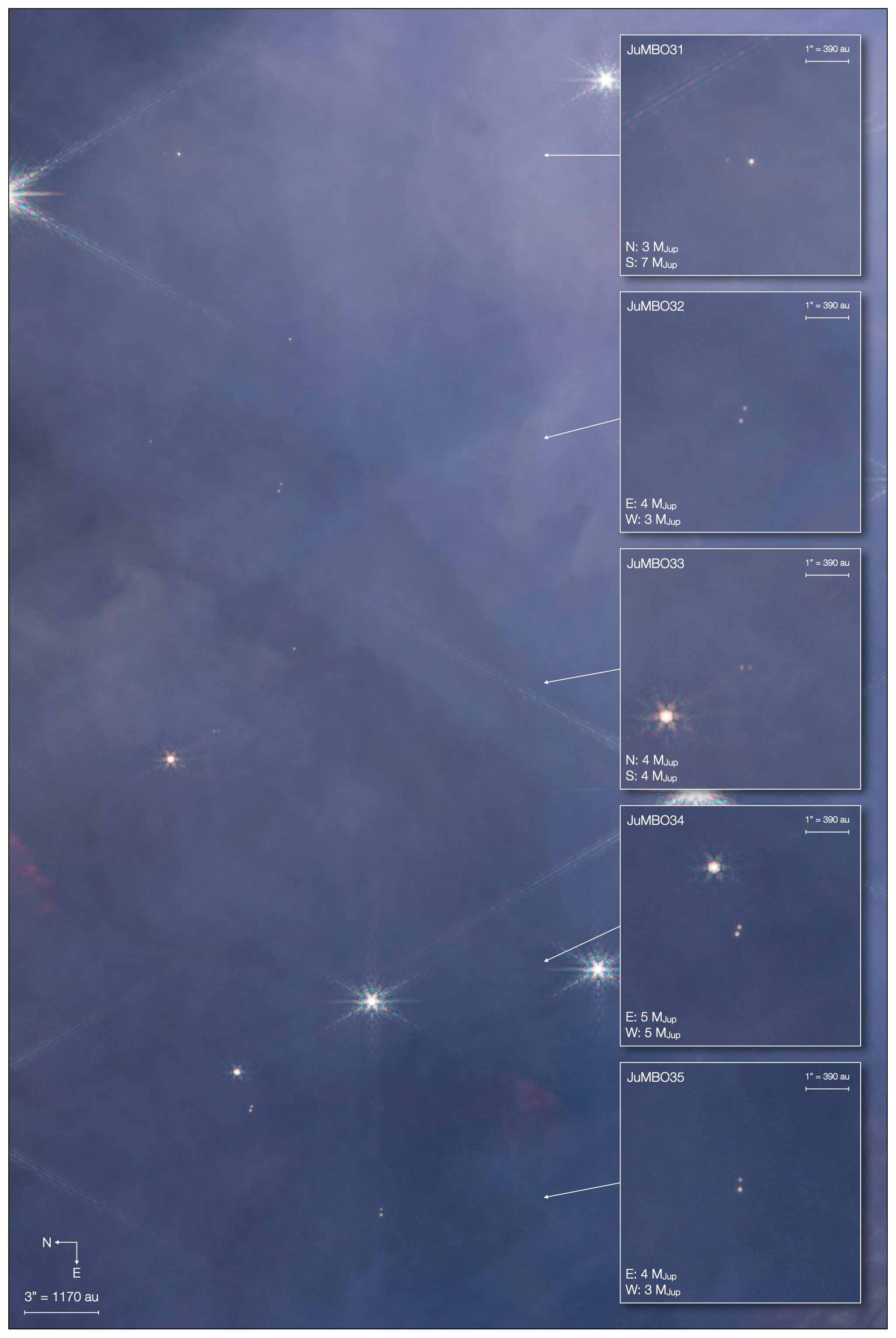}
\caption{A subsection of the full JWST NIRcam short-wavelength colour composite image of the Orion Nebula, located to the east of the Trapezium and south of the Dark Bay. It is centred at 05h35m27.0s, $-05$\degree{}23'27'' (J2000.0) and covers $52.3 \times 35.3$ arcsec or $0.10 \times 0.067$\,pc assuming a distance of 390\,pc. The image has been rotated with N left and E down to show this E-W strip of JuMBOs more effectively. Five JuMBOs are highlighted with zoomed cutouts: all ten of these PMOs have masses $< 7$\Mjup}\label{JUMBOS}
\end{figure}

Figure~\ref{JUMBOS} shows a subsection of the full JWST short-wavelength colour composite mosaic (McCaughrean \& Pearson 2023, submitted), located to the east of the Trapezium and south of the Dark Bay. Five JuMBOs are seen in this one small region, as highlighted in the cutouts. All ten of the individual PMO candidates in these systems are best fit by 1\,Myr evolutionary models with masses 3--7\Mjup{} and thus \Teff{} = 900--1200\,K\@. The \water{} and \methane{} in their SEDs, along with the background molecular cloud, rules out background field stars, and we can also rule out the possibility that these may be foreground T-dwarfs using the known space densities of field brown dwarfs \citep{best2021} combined with the footprint and depth of the NIRCam observations: we would expect to find $<2$ field brown dwarfs across the full NIRCam mosaic \citep{Scholz2022}. Furthermore, as is immediately evident just visually, we can exclude the possibility that many if any of these are chance alignments: based on the  density of sources in our survey, we can calculate that we would expect to find 3.1 chance alignments within 1 arcsec across the whole region. 

Assuming then that the JuMBOs are real binary PMOs, we can compare their statistical properties properties (see Table. \ref{JuMBO_tab}) with higher-mass systems. The JuMBOs span the full mass range of our PMO candidates, from 13\Mjup{} down to 0.7\Mjup. They have evenly distributed separations between $\sim$25--390\,au, which is significantly wider than the average separation of brown dwarf-brown dwarf binaries which peaks at $\sim 4$\,au \citep{close2003, Offner2022}. However, as our imaging survey is only sensitive to visual binaries with separations $> 25$\,au, we can not rule out an additional population of JuMBOs with closer orbits. For this reason we take 9\% as a lower bound for the PMO multiplicity fraction. The average mass ratio of the JuMBOs is $q = 0.66$. While there are a significant number of roughly equal-mass JuMBOs, only 40\% of the them have $q \geq 0.8$. This is much lower than the typical mass ratios for brown dwarfs, which very strongly favour equal masses \citep{close2003, Offner2022}.

\begin{figure}[H]%
\centering
\includegraphics[width=0.9\textwidth]{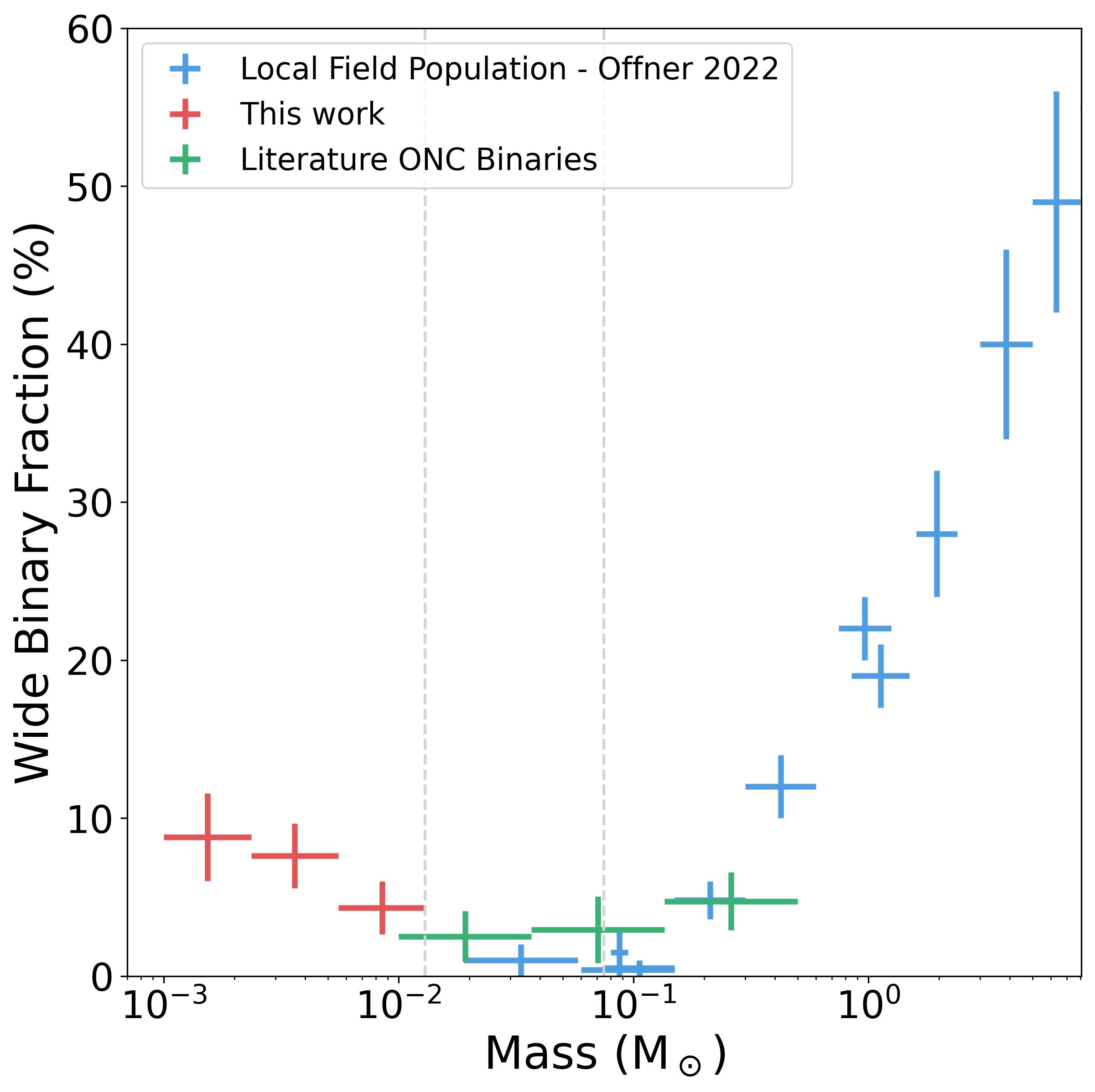}
\caption{The wide binary fraction (WBF) as a function of primary mass, where wide is defined as projected separations $\geq$100\,au, equivalent to 0.26 arcsec at the distance of the Trapezium Cluster. For each point, the horizontal bar indicates the mass interval and the height of the vertical bar indicates the statistical uncertainty in the WBF\@. Blue points show a compilation of multiplicity surveys for the solar neighbourhood \citep{Offner2022}. Green points show the WBF for stars and brown dwarfs in the Trapezium Cluster \citep{Reipurth2007, Correia2013, Kounkel2016, Kounkel2016b, Duchene2018, Jera2019, DeFurio2019, Robberto2020, Strampelli2020, DeFurio2022, DeFurio2022Sub}. Red points are for PMOs in the Trapezium Cluster from this work.}
\label{BinFrac}
\end{figure}

Figure~\ref{BinFrac} shows the wide binary fraction (WBF) as a function of primary mass, where wide is defined as projected separations $\geq$100\,au, equivalent to 0.26 arcsec at the distance of the Trapezium Cluster. Each data point is illustrated with a cross: the horizontal bar indicates the mass interval, while the height of the vertical bar shows the statistical uncertainty in the WBF\@. The blue points show a compilation of multiplicity surveys of the stellar neighbourhood \citep{Offner2022}. The green points show the WBF for stars and brown dwarfs in the Trapezium Cluster calculated by compiling known binaries from the literature \citep{Reipurth2007, Correia2013, Kounkel2016, Kounkel2016b, Duchene2018, Jera2019, DeFurio2019, Robberto2020, Strampelli2020, DeFurio2022, DeFurio2022Sub}. The red points are from this work and show the WBF for PMOs in the Trapezium Cluster. The WBF starts at $50-60$\% for massive stars and decreases monotonically across three orders of magnitude in mass, down to $\sim$2\% in the brown dwarf regime in the Trapezium Cluster. This is consistent with the current predictions of star formation models and the consensus view that the more massive brown dwarfs ($>30{\rm M}_{\rm Jup}$) share the same formation mechanisms as stars \citep{luhman2012, Whitworth2007, Parker2023}. 

However, the PMOs clearly break that trend and prediction, rising back up to at least 9\%. The sudden divergence and increased WBF at planetary masses suggests that new formation mechanisms must come into play at such masses. Broadly speaking, there are two key formation scenarios to consider. If the JuMBOs formed via a ``star-like'' mechanism, \idest{} via core collapse and turbulent fragmentation, then there must be some fundamental extra ingredient involved at these very low masses. Indeed, the JuMBOs in our sample cover the whole range of PMO masses, down to 0.7\Mjup, well below the minimum mass that is thought to be able to form via 3D fragmentation or 2D shocks \citep{Low1976, Rees1976, boyd2005, Whitworth2007}: the formation of such low mass objects raises significant questions in itself.  

Alternatively, perhaps the JuMBOs formed through a ``planet-like'' mechanism in a circumstellar disk around a host star and were violently ejected. Ejections can be caused through planet-planet scattering in the disk \citep{Smullen2016} or by dynamical interactions between stars \citep{bonnell2001}. The latter are relatively common in dense star-forming regions like the Trapezium Cluster. In either case, however, how {\em pairs\/} of young planets can be ejected simultaneously and remain bound, albeit weakly at relatively wide separations, remains quite unclear. The ensemble of PMOs and JuMBOs that we see in the Trapezium Cluster might arise from a mix of both of these ``classical'' scenarios, even if both have significant caveats, or perhaps a new, quite separate formation mechanism, such as a fragmentation of a star-less disk is required \citep{Bodenheimer1978, Tohline2002}.

The advent of JWST marks an exciting milestone for the field of star and planet formation, where observations of isolated objects down to and below 1\Mjup{} will soon become routine. Imaging with NIRCam will reveal many candidates through filter-based SEDs as shown here, while follow-up spectroscopy with NIRSpec (and in lower-density regions with NIRISS) will allow us to place much tighter constraints on their effective temperatures, spectral types, and chemical compositions. 
We will obtain NIRSpec prism spectra of many of the Trapezium Cluster PMO and JuMBO candidates as part of JWST programme 2770 in spring 2024. It would be particularly beneficial to see how the demographics of PMOs change as a function of environmental parameters such as cluster density and how they evolve with age, as this may provide crucial insights that allow us to differentiate between the ``star-like'' and ``planet-like'' formation scenarios for PMOs and JuMBOs alike. It is also clear that further simulations and modelling will be needed to understand how a substantial population of objects can form below 5\Mjup{} and how a significant fraction of them can end up in multiple systems.

\pagebreak
\section{Supplementary Materials}

\subsection{Observations}
The data presented in this paper were obtained with the Near Infrared Camera (NIRCam) onboard the James Webb Space Telescope (JWST), as part of Cycle 1 GTO programme 1256, (P.I. M. McCaughrean). The observations cover an $11' \times 7.5'$ area focused on the inner region of the Trapezium Cluster. A total of 34.9 hours of observing were carried out between September 26th and October 2nd 2022, and split between 12 filters: F115W, F140M, F162M, F182M, F187N, F212N, F277W, F300M, F335M, F360M, F444W and F470N. The observations are split into two mosaic patterns. For two wide-band filters (F115W and F444W), a $7 \time 2$ mosaic covers the $11' \times 7.5'$ field with considerable overlap between the rows and columns. Due to the mosaic pattern, the exposure time is not uniform across the field. This was chosen to ensure accurate registration of the full mosaic using stars in the overlapping regions to yield a good astrometric base for future proper motion studies. These observations used the INTRAMODULEX dither pattern, with four primary dithers, the BRIGHT1 (NGROUPS = 6, NINT = 1) readout pattern and a total exposure time of 515 seconds per visit. This readout pattern was selected to maximise the dynamic range. As the Trapezium Cluster contains a significant number of bright, massive stars, these sources will inevitably be saturated. However, maximising the dynamic range ensures a solid overlap between JWST and existent ground and space-based photometry for intermediate-bright sources in order to bootstrap calibrate the faintest sources in the JWST data. For the remaining five pairs of filters, a $5 \times 2$ mosaic covers the same region but with only marginal overlap in rows, allowing for a more efficient use of observing time. These observations used the INTRAMODULEX dither pattern with six primary dithers, the SHALLOW2 (NGROUPS = 3, NINT = 1) readout pattern to reduce data volumes and a total exposure time of 773 seconds per visit. For further details on the observations, JWST and its instruments and the Trapezium Cluster, we direct the reader to Paper I. Observations \& overview (McCaughrean \& Pearson 2023).

\subsection{Data reduction}
To reduce the observations, we retrieved the stage 0 data products from the Barbara A. Mikulski Archive for Space Telescopes (MAST) and re-ran the stage 1, 2 and 3 reduction steps using a custom version of the 1.11.3 pipeline and Calibration Reference Data System mapping jwst \emph{pmap\_1100}. Stage 1 was run using the optional step argument ${\rm det1.ramp\_fit.suppress\_one\_group = False}$. Stage 2 was run using the default reduction pipeline. A custom version of the stage 3 pipeline was used to align the individual images to Gaia Data Release 3 (GDR3) \citep{Gaia1, Gaia2} and combine the images into the final full mosaics. A brief summary of this process is given below.

The WCS of visit 2 for (F140M, F162M, F182M, F187N, F212N, F277W, F300M, F335M and F360M) and visit 7 for (F115W and F444W) were found to be offset by $\sim$15 arcseconds. This was corrected by manually adding an offset to the wcs data stored in data model in the asdf tree in the fits header of each \_cal.fits file. This is was an approximate correction that does not take into account distortion effects, but significantly reduced the search radius needed for later fine alignment.

We first aligned the F470N data to GDR3, as this filter had the largest overlap between the faintest Gaia sources and unsaturated JWST sources. We compiled an absolute reference catalog of $\sim650$ high quality GDR3 sources that excluded flagged binaries, close pairs, extended galaxies and knots of nebulosity; the latter being a large source of contamination for \HII{} regions such as the Trapezium Cluster. As this catalogue forms the basis of the alignment, care should be taken to remove spurious sources in order to achieve an accurate registration.

For each of the stage 2 \_cal.fits images we compiled an individual source catalogue. The x, y coordinates of the centre of the corresponding GDR3 sources were determined using a non-pipeline recentring routine. Each source was also weighted depending on the quality of the fit and whether it was found to be saturated in the \_cal.fits data. The stage 3 TweakReg routine was then run on each of the \_cal.fits individually. The absolute reference catalogue was passed to the TweakReg routine using the ${\rm tweakreg.abs\_refcat = path\_to\_file}$ step argument. The source catalogues were saved as .ecsv files and were passed to the TweakReg routine by updating the asn with the file path. This process was repeated for each \_cal.fits file individually, as the pipeline defaults to expanding the absolute reference catalogue, which causes alignment errors. The individually aligned files were then resampled into a full combined mosaic using step arguments: ${\rm tweakreg.skip = True}$, ${\rm skymatch.skip = True}$, ${\rm resample.fillval = `nan'}$.

From this F470N mosaic a new absolute reference catalog of $\sim$1500 sources was constructed. The F470N absolute reference catalogue had significantly more overlap of non-saturated sources than GDR3 for the remaining filters, which improved the alignment. This catalogue was used to repeat the above process for the remaining 11 filters, aligning the individual \_cal.fits files to the F470N absolute reference catalogue and then combining and resampling the full mosaics.

\subsection{Source detection}
Sources were detected in the level 3 mosaics produced by stage 3 of the pipeline. First, the two dimensional background of each image was estimated and subtracted using the DAOPHOT MMM algorithm as implemented in Astropy \citep{bradley2023, astropy:2022} using a $30 \times 30$ pixel box and a $5 \times 5$ pixel filter. We used the MMMBackground algorithm to divide the input data into a grid of $30 \times 30$ pixels boxes and then used its mode estimator of the form (3 $\times$ median) - (2 $\times$ mean) to calculate the background level of each box, thus creating a low resolution background map. This image was then median filtered to suppress local under or over estimations, with a window of size of $5 \times 5$ pixels. The final background map was calculated by interpolating the low-resolution background map. Sources were then identified using DAOStarFinder with a threshold of $2\sigma$ and a model PSF for each of the 12 JWST filters employed \citep{Perrin2014}. Sources that were detected in $\geq$ 3 filters were then added to a preliminary source catalogue, which was checked by eye against the images to remove spurious sources, such as bad pixels, knots of nebulosity, diffraction spikes, and persistence spots that had been erroneously flagged as point sources. The by-eye examination was also used to visually classify other sources including proplyds, outflows, and galaxies. The final catalogue contains 3092 sources.

\subsection{Aperture photometry}
Aperture photometry was performed using Photutils \citep{bradley2023}, a package of Astropy \citep{astropy:2022}. We used the aperture\_photometry routine to obtain fluxes for all of the sources in our catalogue, using apertures of 2.5 and 4.5 pixels radius for the sources, while the background was measured in an annulus with inner and outer radii of 5 and 10 pixels, respectively, using a sigma-clipped median. The PIXAR\_SR header keyword was used to convert from surface brightness (${\rm MJy\ sr^{-1}}$) to point source flux (Jy) and then to Vega magnitudes using the zero-points provided by the Spanish Virtual Observatory (SVO) filter profile service \citep{Rodrigo2020}. To convert the aperture magnitudes to total magnitudes, we used the aperture corrections provided by the JWST reference files for the respective filter, interpolated to the corresponding aperture radius. 

\subsection{Extended sources}
Extended sources were identified using aperture photometry and comparing the apparent magnitudes that are calculated with inner apertures of 2.5 and 4.5 pixels. Unresolved point sources will have the same apparent magnitude independent of the choice of inner aperture, whereas extended sources, such as background galaxies and nebular knots, will appear brighter with larger apertures. Sources where the median difference across all 12 filters between 2.5 and 4.5 pixel apertures was greater than 0.1 mag were classified as extended. Sources with a neighbour within 1 arcsec were excluded from this automated classification and checked manually. As well as galaxies, this selection has the potential to flag highly embedded objects, objects with resolved disks and outflows, unresolved binaries and objects in highly featured areas of gas and dust. For this reason our sample of young PMO candidates may not be fully complete, but will be a clean sample of reliable candidates.

\subsection{Evolutionary models}
Throughout our analysis we have utilised the CBPD22 evolutionary models \citep{Chabrier2023}, which combine the atmospheric models from ATMO 2020 \cite{Philips2020} with the a new equation of state for dense hydrogen-helium mixtures \citep{Chabrier2021}. In earlier models the equation of state was based on the so-called additive volume law \citep{Saumon1995, Chabrier2019}, which does not take into account the interactions between hydrogen and helium species. This updated equation of state takes these interactions into account and modifies the thermodynamic properties of the H/He mixture. This primarily affects the entropy profiles, which in turn alters the development of degeneracy and internal structure. The ATMO 2020 models use a 1D radiative-convective equilibrium code to generate three grids of atmospheric models, one using equilibrium chemistry (CEQ) and two using non-equilibrium chemistry (NEQ\_weak \& NEQ\_strong). The non-equilibrium models use a \textit{weak} and a \textit{strong} scaling relation for the eddy diffusion coefficient with surface gravity, which alter the vertical mixing relationships. These models cover the planetary mass regime from $(0.0004{\rm M}_{\odot} - 0.015{\rm M}_{\odot})$. In order to cover the brown dwarf and stellar mass ranges $(0.01{\rm M}_{\odot} - 1.4{\rm M}_{\odot})$ we have used the BHAC15 evolutionary models \citep{Baraffe2015}.

\pagebreak
\section{JuMBO Catalogue}
\begin{table}[!ht]
    \label{JuMBO_tab}
    \centering
    \begin{tabular}{cccccccccc}
        \hline
        Name & RA (deg) & DEC (deg) & M\_Pri & Av\_Pri & M\_Sec & Av\_Sec & Proj\_Sep & M\_Ter & Av\_Ter \\ \hline
        JuMBO 1 & 83.716375 & -5.374688 & 0.001 & 6.3 & 0.001 & 4.3 & 357.7 & - & - \\ 
        JuMBO 2 & 83.718439 & -5.391585 & 0.002 & 16.4 & 0.002 & 13.1 & 114.7 & - & - \\ 
        JuMBO 3 & 83.720854 & -5.379591 & 0.003 & 19.7 & 0.003 & 10.8 & 52.3 & - & - \\ 
        JuMBO 4 & 83.727380 & -5.444921 & 0.002 & 23.7 & 0.001 & 10.6 & 324.4 & - & - \\ 
        JuMBO 5 & 83.727997 & -5.389459 & 0.003 & 10 & 0.002 & 32.8 & 384.3 & - & - \\ 
        JuMBO 6 & 83.734156 & -5.368803 & 0.003 & 46.6 & 0.003 & 56.5 & 70.2 & - & - \\ 
        JuMBO 7 & 83.735012 & -5.387694 & 0.001 & 17.4 & 0.001 & 17.3 & 119 & - & - \\ 
        JuMBO 8 & 83.736001 & -5.445662 & 0.002 & 21 & 0.002 & 15.9 & 101.2 & - & - \\ 
        JuMBO 9 & 83.736884 & -5.332175 & 0.001 & 13.1 & 0.0007 & 8.8 & 211.5 & - & - \\ 
        JuMBO 10 & 83.748149 & -5.445690 & 0.001 & 6.9 & 0.001 & 8.9 & 342.5 & - & - \\ 
        JuMBO 11 & 83.753378 & -5.431788 & 0.0008 & 10.4 & 0.0007 & 15.9 & 192.2 & - & - \\ 
        JuMBO 12 & 83.753580 & -5.354639 & 0.003 & 20.1 & 0.001 & 19.8 & 366.2 & - & - \\ 
        JuMBO 13 & 83.760064 & -5.393619 & 0.001 & 20.5 & 0.001 & 26.5 & 192.6 & - & - \\ 
        JuMBO 14 & 83.767052 & -5.406016 & 0.009 & 39.5 & 0.008 & 36 & 55.6 & - & - \\ 
        JuMBO 15 & 83.768695 & -5.440258 & 0.003 & 39.8 & 0.002 & 26.5 & 329.8 & - & - \\ 
        JuMBO 16 & 83.769429 & -5.415209 & 0.001 & 5.3 & 0.001 & 6.5 & 273.9 & - & - \\ 
        JuMBO 17 & 83.775698 & -5.432976 & 0.001 & 24.5 & 0.0006 & 10.7 & 194.9 & - & - \\ 
        JuMBO 18 & 83.779749 & -5.424113 & 0.003 & 11.7 & 0.002 & 6.6 & 150.6 & - & - \\ 
        JuMBO 19 & 83.785686 & -5.345893 & 0.003 & 22.6 & 0.002 & 31.5 & 273.6 & - & - \\ 
        JuMBO 20 & 83.786364 & -5.411568 & 0.003 & 19.1 & 0.002 & 11.3 & 149.4 & - & - \\ 
        JuMBO 21 & 83.788762 & -5.398635 & 0.007 & 74.2 & 0.002 & 26.1 & 200.5 & - & - \\ 
        JuMBO 22 & 83.801462 & -5.342754 & 0.004 & 51.6 & 0.003 & 29.4 & 127.4 & - & - \\ 
        JuMBO 23 & 83.829058 & -5.446920 & 0.004 & 35.2 & 0.002 & 11.3 & 314.7 & - & - \\ 
        JuMBO 24 & 83.831262 & -5.394369 & 0.011 & 3.6 & 0.011 & 3.5 & 28 & - & - \\ 
        JuMBO 25 & 83.836455 & -5.371124 & 0.005 & 14.2 & 0.004 & 16.4 & 46.1 & 0.004 & 6.1 \\ 
        JuMBO 26 & 83.838007 & -5.366544 & 0.008 & 12.5 & 0.003 & 9.1 & 267.1 & - & - \\ 
        JuMBO 27 & 83.846621 & -5.399533 & 0.009 & 2.4 & 0.002 & 2.8 & 333.1 & - & - \\ 
        JuMBO 28 & 83.846940 & -5.392726 & 0.011 & 8.7 & 0.009 & 20.1 & 58.9 & - & - \\ 
        JuMBO 29 & 83.847252 & -5.346677 & 0.012 & 11.9 & 0.003 & 14.4 & 135 & - & - \\ 
        JuMBO 30 & 83.848540 & -5.405963 & 0.005 & 33.1 & 0.002 & 2.2 & 374.1 & - & - \\ 
        JuMBO 31 & 83.856732 & -5.387897 & 0.007 & 12.8 & 0.003 & 15.2 & 206.7 & - & - \\ 
        JuMBO 32 & 83.860453 & -5.388966 & 0.004 & 14.4 & 0.003 & 11.9 & 118 & - & - \\ 
        JuMBO 33 & 83.863086 & -5.388234 & 0.004 & 17.8 & 0.004 & 23.1 & 73.7 & - & - \\ 
        JuMBO 34 & 83.867221 & -5.388611 & 0.005 & 15.4 & 0.005 & 13.9 & 66.4 & - & - \\ 
        JuMBO 35 & 83.868427 & -5.390019 & 0.004 & 10.1 & 0.003 & 10.3 & 84.5 & - & - \\ 
        JuMBO 36 & 83.878803 & -5.340274 & 0.013 & 32.3 & 0.004 & 36 & 363 & - & - \\ 
        JuMBO 37 & 83.882254 & -5.330745 & 0.003 & 18.3 & 0.002 & 32.2 & 317.6 & - & - \\ 
        JuMBO 38 & 83.883267 & -5.351932 & 0.004 & 27.8 & 0.002 & 24.4 & 213.6 & - & - \\ 
        JuMBO 39 & 83.886789 & -5.372932 & 0.004 & 41.9 & 0.002 & 32.9 & 251 & - & - \\ 
        JuMBO 40 & 83.886856 & -5.364031 & 0.005 & 18.1 & 0.005 & 23 & 164.3 & - & - \\ 
        JuMBO 41 & 83.887251 & -5.375283 & 0.011 & 31.7 & 0.0008 & 17.2 & 287.2 & - & - \\ 
        JuMBO 42 & 83.897548 & -5.333713 & 0.003 & 17.8 & 0.0007 & 15.2 & 123.3 & 0.0007 & 10.8 \\ 
        \hline
        \caption{A short summary of the key JuMBO properties. All masses are in units of \Msolar, projected separation are given in au. For an extend version of this table that includes photometry, see the supplementary catalogue.}
    \end{tabular}
\end{table}

\pagebreak
\section{Data Availability}
The data presented in this paper were obtained with the Near Infrared Camera (NIRCam) onboard the NASA/ESA/CSA James Webb Space Telescope (JWST), as part of Cycle 1 GTO programme 1256, (P.I. M. McCaughrean). They are available on the Barbara A. Mikulski Archive for Space Telescopes (MAST): \url{http://dx.doi.org/10.17909/vjys-x251}.


\section{Acknowledgements}
SGP acknowledges support through the ESA research fellowship programme. The time used to make these JWST observations come from the Guaranteed Time Observation allocation made to MJM upon selection as one of two ESA Interdisciplinary Scientists on the JWST Science Working Group (SWG) in response to NASA AO-01-OSS-05 issued in 2001. SGP would like to thank Victor See for helpful discussions and Katja Fahrion for valuable insights on the JWST calibration pipeline. This work has made use of data from the European Space Agency (ESA) mission {\it Gaia} (\url{https://www.cosmos.esa.int/gaia}), processed by the {\it Gaia} Data Processing and Analysis Consortium (DPAC, \url{https://www.cosmos.esa.int/web/gaia/dpac/consortium}). Funding for the DPAC has been provided by national institutions, in particular the institutions participating in the {\it Gaia} Multilateral Agreement. This research has made use of the Spanish Virtual Observatory (\url{https://svo.cab.inta-csic.es}) project funded by MCIN/AEI/10.13039/501100011033/ through grant PID2020-112949GB-I00. This work made use of Astropy:\footnote{http://www.astropy.org} a community-developed core Python package and an ecosystem of tools and resources for astronomy \citep{astropy:2022}. This research made use of Photutils, an Astropy package for detection and photometry of astronomical sources \citep{bradley2023}.









\bibliography{sn-bibliography}

\end{document}